\documentclass[aps,prb,twocolumn,showpacs,superscriptaddress,amsmath,amssymb]{revtex4-1}
\usepackage{graphicx}
\usepackage{dcolumn}
\usepackage{bm}

\usepackage{multirow}
\usepackage{epsfig}
\usepackage{amssymb}
\usepackage{color}
\usepackage{slashed}
\usepackage{braket}

\begin{document}
\title{Nonvolatile spin field effect transistor based on VSi$_2$N$_4$/Sc$_2$CO$_2$ multiferroic heterostructure}

\author{Xian Zhang}
\affiliation{College of Mechanical and Materials Engineering, Xi'an University, Xi'an 710065, China}

\author{Bang Liu}
\affiliation{State Key Laboratory for Mechanical Behavior of Materials and School of Materials Science and Engineering, Xi'an Jiaotong University, Xi'an, Shaanxi, 710049, China}

\author{Junsheng Huang}
\affiliation{State Key Laboratory for Mechanical Behavior of Materials and School of Materials Science and Engineering, Xi'an Jiaotong University, Xi'an, Shaanxi, 710049, China}

\author{Xinwei Cao}
\affiliation{College of Mechanical and Materials Engineering, Xi'an University, Xi'an 710065, China}

\author{Yunzhe Zhang}
\affiliation{College of Mechanical and Materials Engineering, Xi'an University, Xi'an 710065, China}

\author{Zhi-Xin Guo}
\email{zxguo08@xjtu.edu.cn}
\affiliation{State Key Laboratory for Mechanical Behavior of Materials and School of Materials Science and Engineering, Xi'an Jiaotong University, Xi'an, Shaanxi, 710049, China}

\date{\today}

\begin{abstract}
In this study, we present first-principles calculations that introduce a novel nonvolatile spin field-effect transistor (Spin-FET) utilizing a van der Waals multiferroic heterostructure, specifically  VSi$_2$N$_4$/Sc$_2$CO$_2$. We demonstrate that inverting the ferroelectric polarization in a Sc$_2$CO$_2$ monolayer can effectively modulate the electronic states of a VSi$_2$N$_4$ monolayer, enabling a transition from half-metal to half-semiconductor. This transition significantly alters the electronic transport properties. Furthermore, we construct a Spin-FET device based on this multiferroic heterostructure and observe that the VSi$_2$N$_4$/Sc$_2$CO$_2$-based Spin-FET exhibits exceptional all-electric-controlled performance. Notably, the inversion of the Sc$_2$CO$_2$ ferroelectric polarization yields a substantial on-off current ratio, approximately 650\%, under a minimal bias voltage of 0.02 V. Additionally, we identify a unique spatially-separated spin-polarized transport phenomenon, wherein pure spin-up electrons transport exclusively through VSi$_2$N$_4$, and spin-down electrons through Sc$_2$CO$_2$. Our findings suggest a promising pathway for developing low-energy-dissipation and nonvolatile FET devices.

\end{abstract}

\maketitle

\section{Introduction}

During the continuous miniaturization of electronic products, Joule heating and quantum limit on devices become the bottlenecks that prevent increases in operation speed and information density, which invalidate the Moore's law \cite{Liu,Xie}. Spintronics is believed to be the strong candidate of the future technology in the post-Si-era \cite{Treger,Marun, ChenPRB2020,Zhang1,Zhang2,NSR2021}. For any operation in spintronics, the techniques for injection, detection, manipulation, transport, and storage of spins need to be established \cite{Nitta}.  In spin-based devices, encoding and reading out spin information in single spins can be considered the ultimate limit for scaling magnetic information. Spin field-effect transistor (Spin-FET) is a fundamental spin-based device for the spin operation \cite{Datta, Datta1}.
Developing low-power cost Spin-FET that can efficiently realize the  \emph{on} or \emph{off} state of pure spin current is crucial for the development of spintronics \cite{Dieny}. 

Traditionally, Spin-FET is a three terminal device with spin polarized current flowing between drain and source terminals and the gate terminal is used to control this current. As proposed by Datta and Das, the spin precession motion of electron can be controlled  through the Rashba spin-orbit coupling effect, where \emph{on} and \emph{off} states are distinguished by $\pi$ phase difference in spin precession motion \cite{Datta, Datta1}. Nonetheless, since the realization of $\pi$ phase difference in a Spin-FET depends on the precise control of gate voltage,  traditional Spin-FETs  meet challenge of the nonvolatile functionality, which is usually realized by using a ferroelectric material as gating.
 
In this work, based on the first-principles calculations, we propose a Spin-FET possessing remarkable nonvolatile  \emph{on} and \emph{off} functionality.  Such Spin-FET uses a two-dimensional (2D) van der Waals (vdW) multiferroic heterostructure (e.g. composed by monolayer VSi$_2$N$_4$ and monolayer Sc$_2$CO$_2$) as the channel material, instead of the nonmagnetic semiconductors (like InAs, InAlAs, etc.) that are generally adopted in traditional Spin-FETs \cite{Parveen}. In the multiferroic heterostructure, ferromagnetic material  VSi$_2$N$_4$  has a half-semiconductor characteristic and is mainly responsible for the spin  transport \cite{Lake,Li}. Whereas, the ferroelectric material Sc$_2$CO$_2$ has a semiconductor characteristic \cite{Yang, Ang} and is mainly in charge of controlling the spin-polarized electronic structure of VSi$_2$N$_4$.  

It is noteworthy that constructing vdW multiferroics with 2D ferroelectrics and ferromagnets and studying their full-electric-control of nonvolatile device applications had been recently explored. Nonetheless, these studies are mostly on the stage of proving a proof-of-concept of all-electric-controlled valving effects \cite{Ang, ZhaoX} accompanied by some pioneer works on the magnetic tunnel junction (MTJ) device applications \cite{WangJ,Lu1,Tsymbal}. The concept of using 2D multiferroic heterostructure to realize the nonvolatile Spin-FETs is first proposed here. Moreover, the all-electric-controlled nonvolatile half-semiconducting to half-metallic transition is still to be realized in 2D ferroelectrics.
 
We find that the inversion of ferroelectric polarization of  Sc$_2$CO$_2$ can efficiently change the spin-up electronic structure of VSi$_2$N$_4$, leading to a half-semiconductor to half-metal transition. 
Moreover,  the VSi$_2$N$_4$/Sc$_2$CO$_2$ based Spin-FET has remarkable  all-electric-controlled performance, where a large on-off current ratio can be obtained under a small bias voltage. Additionally, an interesting spatially-separated spin-polarized transport phenomenon, i.e.,  pure spin-up (spin-down) current transporting in VSi$_2$N$_4$ (Sc$_2$CO$_2$), is observed.

\section{method}
Geometric optimization and electronic structures of the heterostructure are calculated by density-functional theory (DFT) with the projector-augmented-wave (PAW) method, which is implemented in the Vienna ab initio simulation package \cite{Vasp1,Vasp2}. The exchange-correlation interaction is treated by the generalized gradient approximation (GGA) based on the Perdew-Burke-Ernzerhof (PBE) function and HSE06 functional for the geometric optimization and electronic structures, respectively \cite{Guo1,Guo2}. The convergence standards of the atomic energy and positions are less than 1 $\times$ 10$^{-5}$ eV per atom and 1 $\times$ 10$^{-2}$ eV \AA$^{-1}$, respectively. The cutoff energy of wave function is set to 500 eV. The Brillouin zone (BZ) is sampled by a 9 $\times$ 9 $\times$ 1 Monkhorst-Pack k-point mesh. 

The transport properties are simulated based on the DFT method combined with the nonequilibrium Green's function (NEGF) formalism, using the Atomistix ToolKit (ATK) 2019 package \cite{ATK}. In the calculations, the Tier 3 basis set of the linear combination of atomic orbital (LCAO) is adopted with Hartwigsen-Goedecker-Hutter pseudopotentials and GGA in the form of the PBE function is utilized to represent the exchange and correlation interactions. The transmission calculations are carried out with the PBE pseudopotentials distributed in the QuantumWise package \cite{ATK,ATK1}. A density mesh-cutoff of 75 Hartree is employed, and the BZ is sampled using a $6\times 1 \times 94$ k-point grid in the transport calculations. At a bias-voltage V$_b$, the spin-resolved current ($I_{\sigma}$) can be calculated as follows: 
\begin{equation}
	I_{\sigma}(V_b) = \frac{e}{h}\int_{-\infty}^{+\infty} dE \left[ f(E, \mu_L)-f(E, \mu_R)\right] T_{\sigma}(E,V_b),
\end{equation}
where $f(E, \mu_L)$ and $f(E, \mu_R)$ are the Fermi-Dirac distribution of the left and right electrodes, respectively. $\mu_L$ and $\mu_R$ are the electrochemical  potentials  of the left and right electrodes, respectively, and $T_{\sigma}(E)$ is the transmission probability for an electron at energy E with spin $\sigma$. Note that the linear response current originated from the equilibrium transmission spectrum at zero bias is considered in the calculations \cite{Guo3}.

\section{Results}
The heterostructure is constructed by a ferromagnetic (FM) monolayer VSi$_2$N$_4$ and a ferroelectric monolayer Sc$_2$CO$_2$. The monolayer Sc$_2$CO$_2$ was predicted to have a sizable out-of-plane electrical polarization as high as 1.60 $\mu$C/cm$^2$, and the reverse of electric polarization is achieved via displacement of  C atoms with an energy barrier about 0.52 eV \cite{Yang}. The Sc$_2$CO$_2$ with carbon-up and carbon-down configurations correspond to the polarization-up (P$\uparrow$) and polarization-down states (P$\uparrow$), respectively (see Fig. 1(a)). On the other hand, VSi$_2$N$_4$ is a theoretically predicted 2D magnetic material, which is a member of the MA$_2$Z$_4$ family. The Curie temperature of monolayer VSi$_2$N$_4$ is predicted to be higher than room temperature, showing the promising potential applications in future spintronic devices \cite{WangL2021,Lake}. Fig. 1(b) further shows the calculated band structure of monolayer  VSi$_2$N$_4$. An interesting characteristic is that there is only spin-up electrons in a wide energy range around the Fermi energy (E$_F$), i.e., [-1.7, 1.4] eV, making  VSi$_2$N$_4$ an ideal half-semiconductor. 

\begin{figure}
\begin{center}
\includegraphics[angle= 0,width=0.8\linewidth]{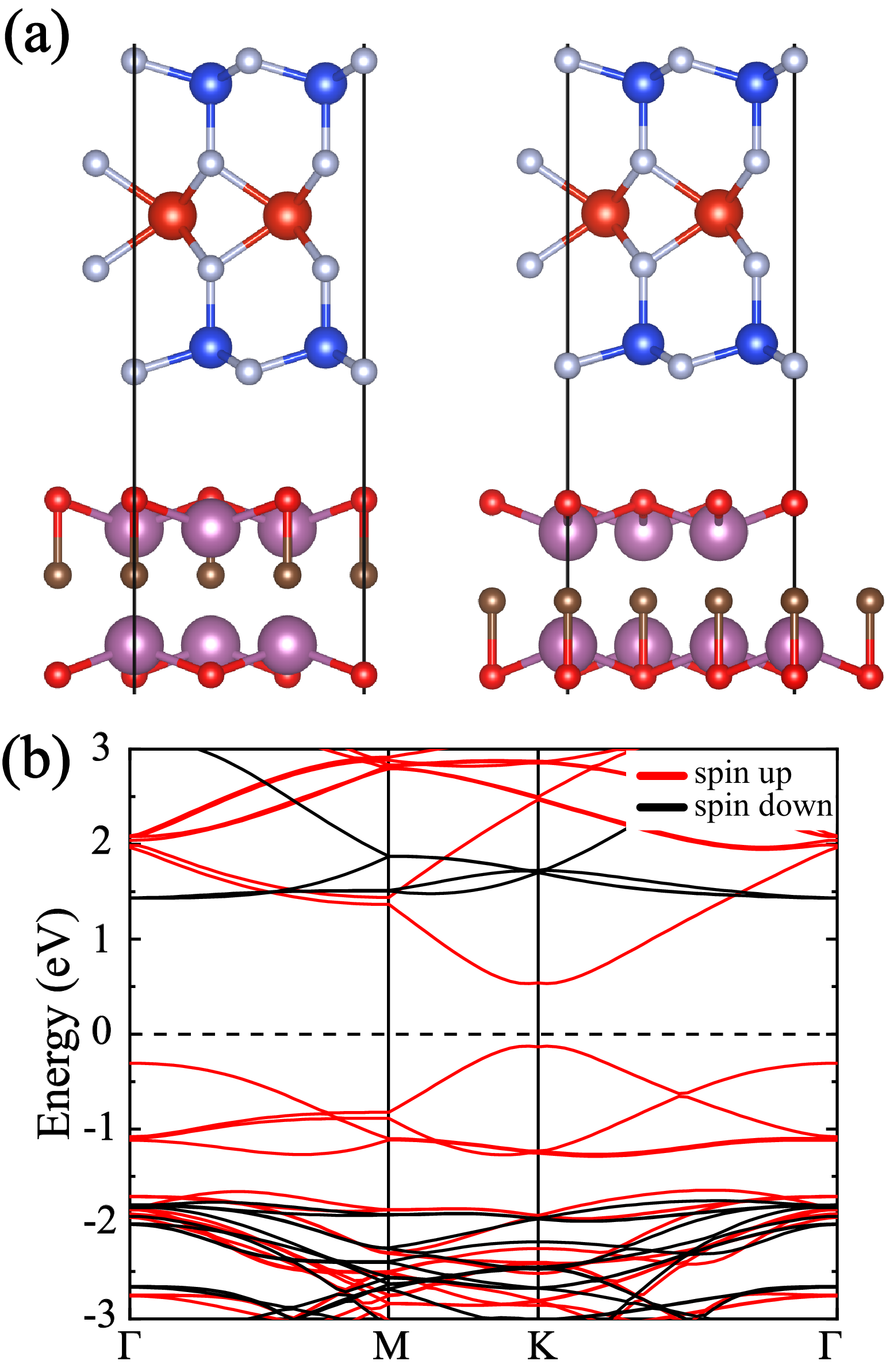}
\caption{
Atomic structures of VSi$_2$N$_4$/Sc$_2$CO$_2$ heterostructure with Sc$_2$CO$_2$  in P$\uparrow$ and P$\downarrow$  polarizations and the spin-resolved energy bands of VSi$_2$N$_4$.  (a) Atomic structures of VSi$_2$N$_4$/Sc$_2$CO$_2$ heterostructure, with Sc$_2$CO$_2$ in P$\uparrow$  (left panel) polarization and P$\downarrow$  (right panel) polarizations, respectively. (b) Band structure of VSi$_2$N$_4$. The grey, large red, and blue balls represent the N, Si, and V atoms, respectively. The red, big purple, and orange  balls represent the O, Sc, and C atoms, respectively.
 }
\label{F1}
\end{center}
\end{figure}

Considering that the in-plane lattice constant of freestanding Sc$_2$CO$_2$ and VSi$_2$N$_4$  are 5.74 \AA~ and 2.88 \AA, respectively, a 2 $\times$ 2  super-periodicity of VSi$_2$N$_4$ is commensurate to the 1 $\times$ 1 Sc$_2$CO$_2$, with a lattice mismatch of 1.23\%.  As shown in Fig. S1, we have considered three typical stacking configurations of the VSi$_2$N$_4$/Sc$_2$CO$_2$ heterostructure, namely, hollow configuration, bridge configuration, and top configuration. {The calculated total energies of the three  configurations are -300776.98 meV, -300776.90 meV, and -300776.88 meV, respectively}, showing that the hollow configuration (also see Fig. 1(a)) is the most stable configuration.  Nevertheless, the total energy difference among different configurations is very small ($\leq$ 0.1 meV), indicating that all the three configurations can be obtained in experiments. The interlayer spacings between VSi$_2$N$_4$ and Sc$_2$CO$_2$  are 2.82   \AA~and 3.06 \AA~for the  P$\uparrow$ and  P$\downarrow$ polarization states, respectively.  It is known that the atomic radii of O and N atoms, which are located at the surface of Sc$_2$CO$_2$ and  VSi$_2$N$_4$, respectively,  are 0.66  \AA~and  0.71 \AA. Hence, the sum of their atomic radii is much smaller than the interlayer spacing, showing the nature of vdW interaction between  Sc$_2$CO$_2$ and  VSi$_2$N$_4$. To ensure  that the VSi$_2$N$_4$/Sc$_2$CO$_2$ heterostructure is energetically stable, we further calculate the binding energy ($E_b$) between VSi$_2$N$_4$ and Sc$_2$CO$_2$, which is defined as $E_b =   (E_{S} + E_{V} - E_{tot})/S$. Here $E_{S}$, $E_{V}$, and $E_{tot}$ are the total energy of monolayer Sc$_2$CO$_2$, monolayer VSi$_2$N$_4$, and VSi$_2$N$_4$/Sc$_2$CO$_2$ heterostructure, respectively. $S$ represents the interface area of the simulated cell. In the hollow configuration, it is found that $E_b$ = -31 meV/\AA$^2$ and  -21 meV/\AA$^2$ for the  P$\uparrow$ and  P$\downarrow$ polarization states, respectively. This result shows that the heterostructure is energetically stable under a vdW interlayer interaction.

Then, we discuss the electronic properties of VSi$_2$N$_4$ manipulated by the polarization of Sc$_2$CO$_2$.
In comparison with the band structure of freestanding VSi$_2$N$_4$ (Fig. 1(b)), there is an obvious upshift of E$_F$ of VSi$_2$N$_4$ in the heterostructure with  Sc$_2$CO$_2$ in the P$\uparrow$ polarization  (Figs. 2(a) and 2(b)). As a result, a half-semiconductor to half-metal transition occurs in VSi$_2$N$_4$. Moreover, the band gap of spin-up electrons remarkably decreases from 0.7 eV to 0.1 eV, showing the strong interface effect on the electronic structure of VSi$_2$N$_4$. It is noticed that the spin-up energy bands in the heterostructure present a type-III band alignment, whereas the spin-down energy bands have a type-II band alignment. This feature indicates that there is an obvious electron transfer from Sc$_2$CO$_2$ to VSi$_2$N$_4$, which fills up the spin-up conduction states. 
As for the case Sc$_2$CO$_2$ in P$\downarrow$ polarization, the spin-up energy bands of VSi$_2$N$_4$  present a type-I band alignment, and the spin-down energy bands form a type-II band alignment (Figs. 2(c) and 2(d)). This result shows that there is little charge transfer between  Sc$_2$CO$_2$ and VSi$_2$N$_4$. Compared to that of freestanding  VSi$_2$N$_4$, there is also a sizable decrease of band gap of the spin-up electrons (from 0.7 eV to 0.2 eV), whereas the band gap of spin-down electrons changes little. This result is similar to that of Sc$_2$CO$_2$ in P$\uparrow$ polarization. It is noticed that the E$_F$ lies in the band gap of spin-up electrons, which makes the heterostructure  a half-semiconductor. 
The above results show that the inversion of out-of-plane polarization of Sc$_2$CO$_2$ can dramatically change the band alignment as well as the band structure of VSi$_2$N$_4$  in the  heterostructure.

\begin{figure}[h]
\begin{center}
\includegraphics[angle= 0,width=0.98\linewidth]{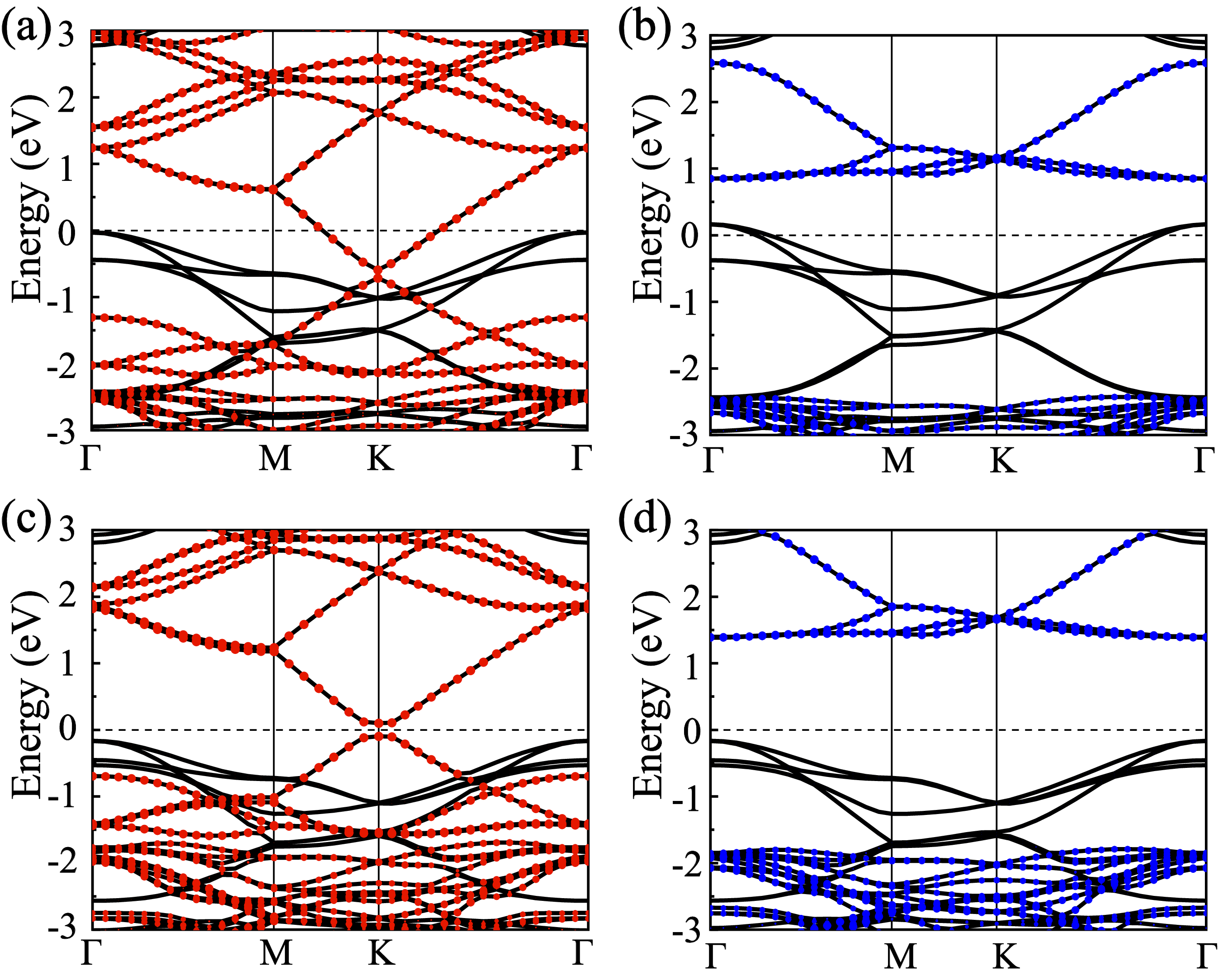}
\caption{
Spin-resolved energy bands of  VSi$_2$N$_4$/Sc$_2$CO$_2$  heterostructure with Sc$_2$CO$_2$ in P$\uparrow$ (a,b) and   P$\downarrow$ (c,d) polarizations. (a) and (c) are the energy bands of spin-up electrons, (b) and (d) are the energy bands of spin-down electrons. Red and blue symbols denote spin-up and spin-down electronic states of VSi$_2$N$_4$, respectively. }
\label{F2}
\end{center}
\end{figure}

To further show the Sc$_2$CO$_2$-polarization dependent charge transfer effect on band structure of VSi$_2$N$_4$, we calculate the electron density difference (EDF) of VSi$_2$N$_4$/Sc$_2$CO$_2$  heterostructure. As shown in Fig. 3(a), in comparison with that of  freestanding VSi$_2$N$_4$ (Sc$_2$CO$_2$),  there are extra electrons (holes) in the VSi$_2$N$_4$ (Sc$_2$CO$_2$) part in the P$\uparrow$ state. According to Bader's charge analysis, around 0.26 e transfers from Sc$_2$CO$_2$ to VSi$_2$N$_4$, resulting in a half-semiconductor to half-metal transition of VSi$_2$N$_4$. On the other hand,  the charge transfer from Sc$_2$CO$_2$ to VSi$_2$N$_4$ in the P$\downarrow$ state is much smaller (0.03 e, see Fig. 3(b)). Hence, VSi$_2$N$_4$ holds a half-semiconductor characteristic. This result agrees well with the Sc$_2$CO$_2$-polarization dependent band alignments in the heterostructure.

To understand the origin of charge transfer from  Sc$_2$CO$_2$ to VSi$_2$N$_4$, we have additionally explored the band alignment between freestanding Sc$_2$CO$_2$ and freestanding VSi$_2$N$_4$. As shown in Fig. S2, type-I band alignment is obtained when  Sc$_2$CO$_2$ is in either P$\uparrow$ or P$\downarrow$ polarizations. This result indicates little charge transfer occurs between Sc$_2$CO$_2$ and VSi$_2$N$_4$ from the viewpoint of relative electrostatic potential. Nonetheless, there is another mechanism for the interfacial charge transfer, i.e., electronegativity difference between elements \cite{Marianetti,Guo4}. In the VSi$_2$N$_4$/Sc$_2$CO$_2$ heterostructure, charge transfer at the interface mainly occurs between N atoms of VSi$_2$N$_4$ (in the bottom layer) and Sc atoms of Sc$_2$CO$_2$ (in the top layer), owing to the large atomic radius (Sc) and short inter-atomic distances.  Since the electronegativity of N and Sc are 3.04 and 1.36, respectively, electrons prefer to transfer from Sc to N, resulting in the charge transfer from Sc$_2$CO$_2$  to VSi$_2$N$_4$. Note that the smaller charge transfer from Sc$_2$CO$_2$ to VSi$_2$N$_4$ in the P$\downarrow$ state  can be attributed to the larger interlayer distance. 

\begin{figure}[h]
	\begin{center}
		\includegraphics[angle= 0,width=0.8\linewidth]{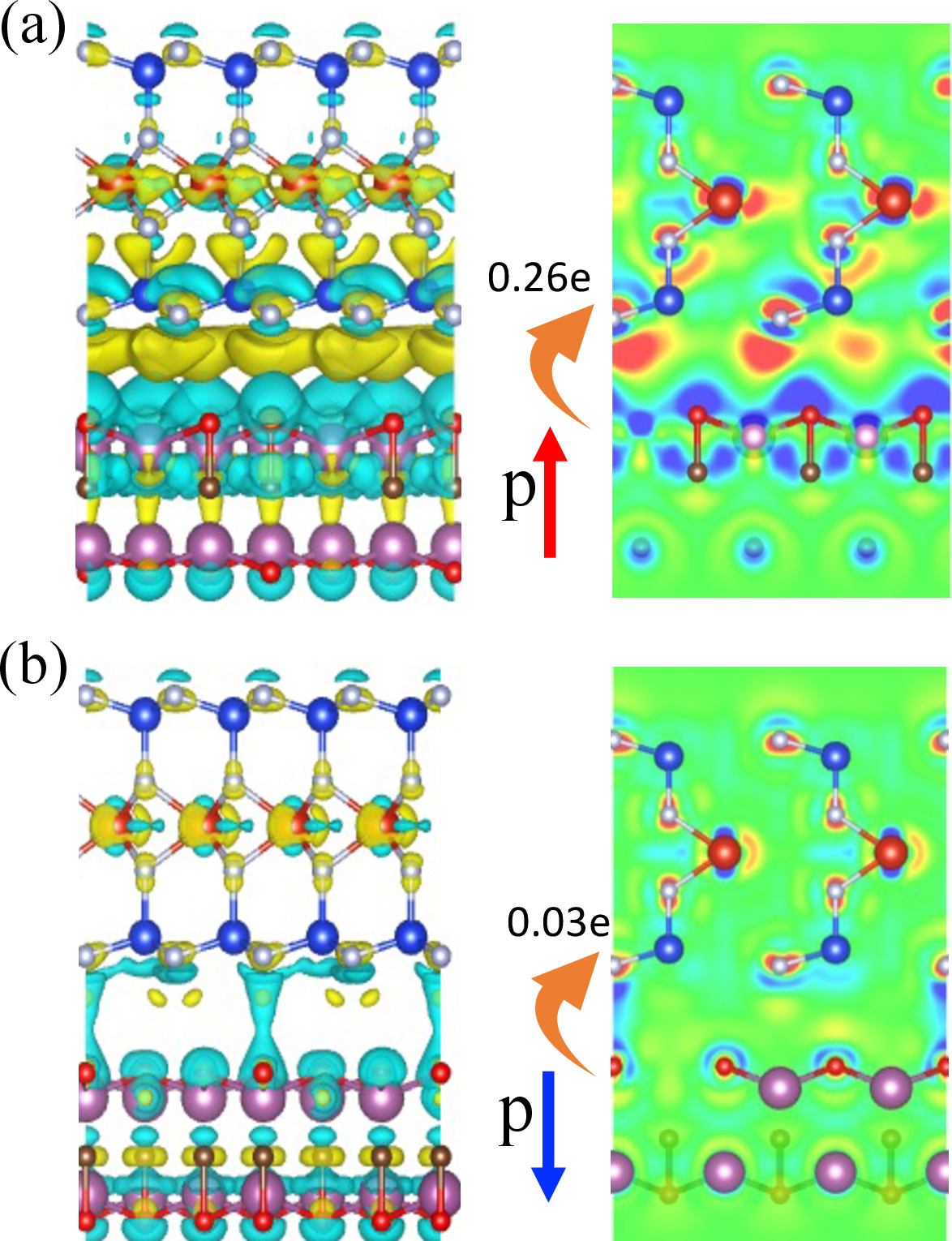}
		\caption{Electron density difference (EDF) of VSi$_2$N$_4$/Sc$_2$CO$_2$ heterostructure. (a) EDF with Sc$_2$CO$_2$ in P$\uparrow$ polarization. (b) EDF with  Sc$_2$CO$_2$ in P$\downarrow$ polarization. The left panels in (a) and (b) are the isosurface of EDF, where the yellow and azure colors represent the positive and negative values, respectively. The right panels in (a) and (b)  are the contour plots of EDF, where the red and blue colors correspond to the normalized maximum of the positive and negative values, respectively.  
		}
		\label{dc1}
	\end{center}
\end{figure}

In addition, the band projection  analysis (Fig. S3) shows that both the  conduction band minimum (CBM) and valence band maximum (VBM) of spin-up electronic states of VSi$_2$N$_4$ are attributed to the d$_{z^2}$ and d$_{x^2-y^2}$ ($e_g$ orbitals of V atoms), which located in the center of VSi$_2$N$_4$. Note that the $e_g$ orbitals of V atoms also give rise to the CBM of spin-down electrons as shown in Fig. S3, and thus induce the type-II band alignment. This result shows that the atomic layer composed of V atoms is mainly in charge of the spin current in the VSi$_2$N$_4$/Sc$_2$CO$_2$ heterostructure.

It is noticed that in both P$\uparrow$ and P$\downarrow$ states, V atoms of  VSi$_2$N$_4$ gain sizable electrons in the heterostructure (Fig. 3). Since both  CBM and VBM are mainly attributed to d orbitals of V, the band gap is particularly sensitive to the electron doping around V atoms, which leads to considerable reduction of band gap. To confirm such an argument, we have additionally calculated the band structure of monolayer  VSi$_2$N$_4$ by electron doping. As shown in Fig. S4, the band gap of VSi$_2$N$_4$ remarkably decreases from 0.7 eV to 0.1 eV with 0.5 e doping in the simulated cell.  It is noteworthy that  VSi$_2$N$_4$ also transforms from a half-semiconductor to a half-metal due to the downshift of CBM, when the doped electron is sizable.

\begin{figure}[h]
\begin{center}
\includegraphics[angle= 0,width=0.98\linewidth]{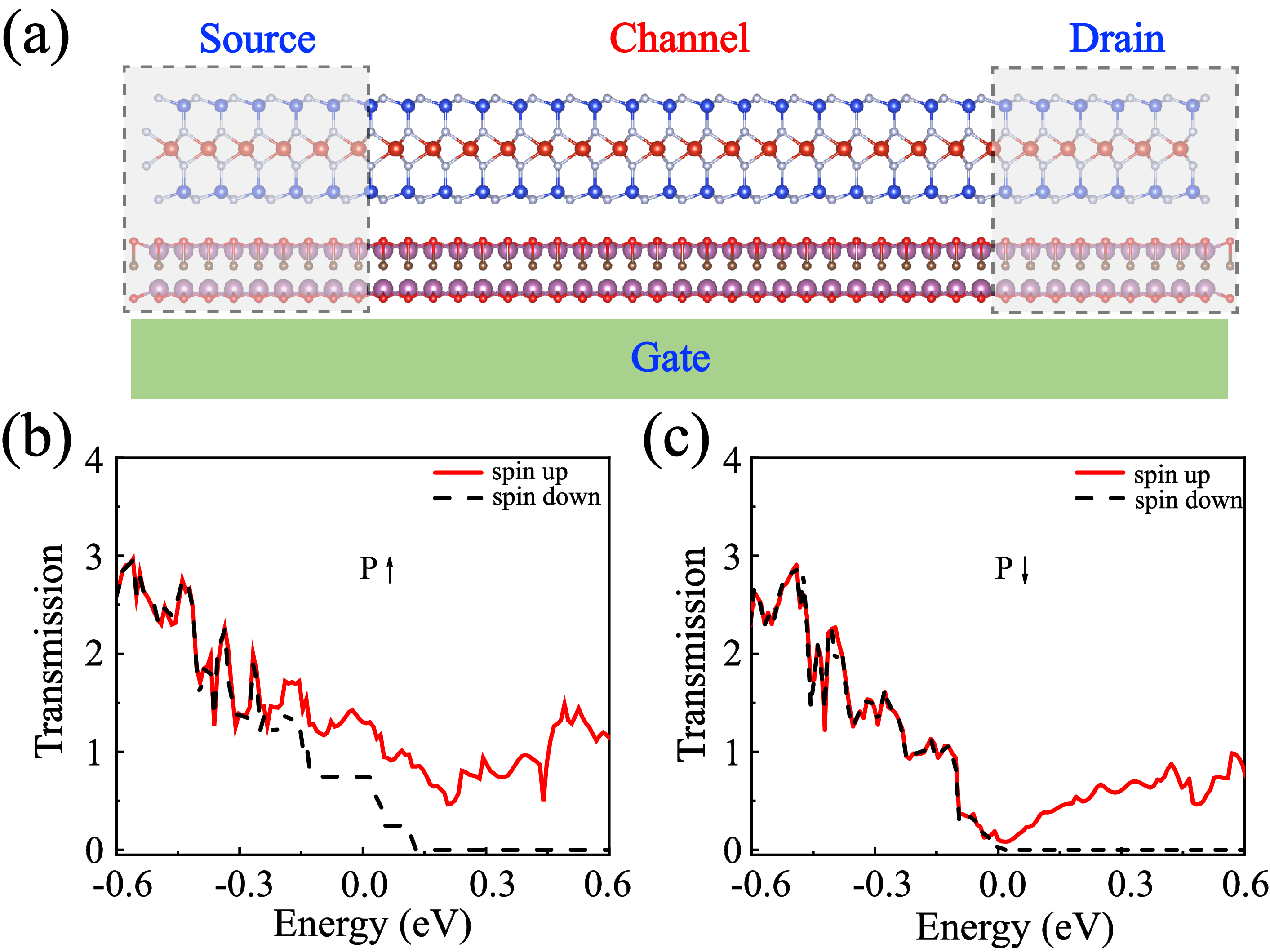}
\caption{ (a) The schematic structure of the proposed Spin-FET based on the VSi$_2$N$_4$/Sc$_2$CO$_2$ heterostructure. (b) The calculated spin-polarized transmission spectrum of the Spin-FET with  Sc$_2$CO$_2$ in P$\uparrow$ polarization. (c) The calculated spin-polarized transmission spectrum of the Spin-FET with  Sc$_2$CO$_2$ in P$\downarrow$ polarization.
  }
\label{dc2}
\end{center}
\end{figure}

Next, we discuss the  plentiful nonvolatile spin transport properties of VSi$_2$N$_4$ induced by  the Sc$_2$CO$_2$ polarizations. We first propose a novel Spin-FET structure based on the VSi$_2$N$_4$/Sc$_2$CO$_2$ heterostructure, as shown in Fig. 4(a). In this structure, one can effectively control the inversion of  Sc$_2$CO$_2$  ferroelectric polarization by a bottom gate. When Sc$_2$CO$_2$ is in  P$\uparrow$ polarization, according to the band structure of VSi$_2$N$_4$/Sc$_2$CO$_2$ shown in Fig. 2(a), one expects a large spin-up current and thus the Spin-FET  is in an \emph{on}  state, due to the half-metal characteristic of  VSi$_2$N$_4$. 
In contrast, when  Sc$_2$CO$_2$ is in P$\downarrow$ polarization, a small spin-up current and thus the \emph{off} state  is expected within a low bias voltage region, owing to the half-semiconductor characteristic of VSi$_2$N$_4$ (Figs. 2(c)). Note that, there is no spin-down  current from VSi$_2$N$_4$, due to the large band gap of the spin-down electrons of VSi$_2$N$_4$ (Figs. 2(b) and 2(d)). Consequently, a pure spin-polarized current can be realized in the VSi$_2$N$_4$/Sc$_2$CO$_2$ based nonvolatile Spin-FET. 

To verify above argument, we have also calculated the electronic transport property of the Spin-FET. Figs. 4(b) and 4(c) show the transmission spectrum of this device at zero bias with  Sc$_2$CO$_2$  in  P$\uparrow$ and  P$\downarrow$ polarizations, respectively. It is seen that the transmission of spin-up electrons in  P$\uparrow$ state reaches up to 1.4 at the Fermi level, whereas it is only 0.1 in the P$\downarrow$ state. The nonzero transmission at Fermi level in the P$\downarrow$ state can be attributed to the  nonzero broadening parameter in the Green's function approach. As a result, an on-off ratio of 14 for the spin-up electrons of VSi$_2$N$_4$ can be obtained in the law bias voltage region. Notably, in the  P$\uparrow$ state, there is also sizable transmission at Fermi level (0.7) for the  spin-down electrons, which comes entirely  from  Sc$_2$CO$_2$ (as shown in Fig. 2(b), the VBM of spin-down electrons of  Sc$_2$CO$_2$ upshifts to 0.2 eV above E$_F$). This result shows an interesting spatially-separated spin-polarized transport phenomenon in the Spin-FET, which has potential applications in the future spintronics. 
Note that the transmission of spin-down electrons of both VSi$_2$N$_4$ and  Sc$_2$CO$_2$ in the  Spin-FET goes to zero at higher energy levels, i.e., E$\ge$0.15 eV and 0.01 eV with Sc$_2$CO$_2$ in  P$\uparrow$ and P$\downarrow$ states, respectively. In this condition, the spin-down current from  Sc$_2$CO$_2$ can be completely inhibited, and thus the pure spin-up current  can be  realized. This result implies the good spin-valve performance of the VSi$_2$N$_4$/Sc$_2$CO$_2$ based devices. In addition, a more distinct transporting difference between the half-metal  VSi$_2$N$_4$ in the heterostructure and the intrinsic  VSi$_2$N$_4$ is expected owing to its larger intrinsic band gap (0.7 eV).

\begin{figure}[h]
\begin{center}
\includegraphics[angle= 0,width=1.0\linewidth]{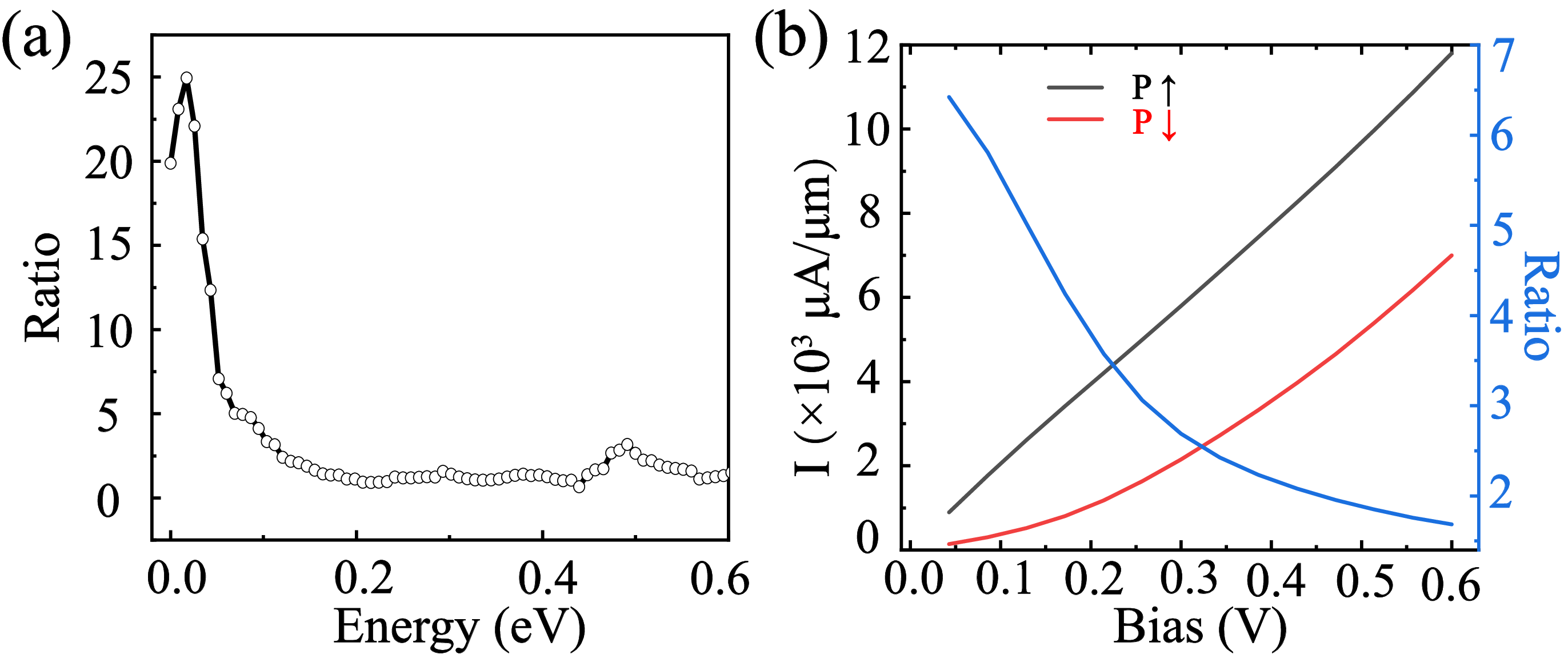}
\caption{ (a) The ratio of total transmission (sum of spin-up and spin-down transmission) between  the P$\uparrow$ and  P$\downarrow$   polarization states. (b) The calculated I-V curves of the total current (sum of spin-up and spin-down current) with Sc$_2$CO$_2$ in P$\uparrow$ (solid red line) and  P$\downarrow$ (solid black line) polarizations, respectively.  The blue line in (b) shows the on-off ratio of total  current.
  }
\label{dc2}
\end{center}
\end{figure}

Then we come to discuss the on-off ratio of the total current in the Spin-FET. Fig. 5(a) shows the on-off ratio of the total transmission (sum of spin-up and spin-down transmissions). It is seen that the on-off ratio monotonically increases with decreasing the bias voltage from 0.6 V, reaching 2500\% around the Fermi level. This result indicates that a drastic total current (sum of spin-up and spin-down currents) difference can be induced  by the inversion  of ferroelectric polarizations of Sc$_2$CO$_2$ in the low bias voltage region. Fig. 5(b) further shows the calculated I-V curves of the total current in P$\uparrow$ and  P$\downarrow$  states, respectively. In the P$\uparrow$ state, the total current linearly increases with the bias voltage. Whereas, in the P$\downarrow$ state it quadratically  increases with the bias voltage.  As a result, the on-off ratio  of  total current monotonically increases with decreasing the voltage, and a large  on-off ratio of about 650\% can be obtained when a small bias voltage (0.02 V) is applied (Fig. 5(b)).  

Finally, we discuss the channel length of the VSi$_2$N$_4$/ Sc$_2$CO$_2$ based spin-FTEs. The channel length can be estimated by the spin diffusion length ($\lambda$), which is inversely proportional to the spin-orbit coupling (SOC) strength of channel materials. Unfortunately, there is no report on $\lambda$ of  VSi$_2$N$_4$ so far. Since N and Si are the light elements in periods 2 and 3, respectively, both of their SOC strength are very weak, contributing to a large $\lambda$. Indeed, it has been found that the bulk Si has a large $\lambda$ of 2-3 $\mu$m \cite{APL2010}. On the other hand, V is a transition metal in period 4, which would be responsible for the lower limit of $\lambda$ in VSi$_2$N$_4$. It has been found that pure V films has a spin diffusion length of about 15 nm due to the strong d-orbital based SOC \cite{PRB2014}. Considering that the d-orbital density of states in VSi$_2$N$_4$ is much smaller than that in V films, one can expect a larger $\lambda$ in VSi$_2$N$_4$, which might be tens of nanometers. Consequently, a channel length about tens of nanometers is expected in the VSi$_2$N$_4$/ Sc$_2$CO$_2$ Spin-FETs.

\section{summary}
In summary, based on the VSi$_2$N$_4$/Sc$_2$CO$_2$ multiferroic heterostructure, we propose a way of realizing nonvolatile Spin-FET, which has a low-energy-consumption characteristic. The DFT calculations show that the inversion of  Sc$_2$CO$_2$ ferroelectric polarization can efficiently modulate the  spin-polarized  electronic structure of VSi$_2$N$_4$, which induces a half-semiconductor to half-metal  transition. 
Such electronic phase transition can lead to significant change of electronic transport properties, which is an essential concept of the nonvolatile Spin-FET. We additionally construct a  Spin-FET device based on the VSi$_2$N$_4$/Sc$_2$CO$_2$  heterostructure. The  electronic transport calculations show that the Spin-FET has remarkable all-electric-controlled performance, where a large on-off ratio (about 650\%) of the total current has been obtained when a small bias voltage (0.02 V) is applied. 
Moreover, an interesting spatially-separated spin-polarized transport phenomenon has been observed when Sc$_2$CO$_2$  is in  P$\uparrow$  polarization, with a pure spin-up (spin-down) current in VSi$_2$N$_4$ (Sc$_2$CO$_2$), respectively.
These findings can be helpful to the design of high-performance nonvolatile 2D Spin-FET devices.

Before closing, we remark that continuously reducing the bias voltage is in line with the requirement of International Roadmap of Semiconductors (ITRS) \cite{ITRS}, which benefits to the low-power chips. Although there has been no specific roadmap for the spintronic chip by far, one can expect a smaller bias to be utilized in spintronic devices due to its low-power nature, which makes the small working bias of VSi$_2$N$_4$/Sc$_2$CO$_2$ based Spin-FET possible. Moreover, a larger bias voltage can be achieved by properly selecting the composite of 2D multiferroic heterostructures, which is to be explored in the future works.

\begin{acknowledgments}
We acknowledge financial support from the Ministry of Science and Technology of the People's Republic of China (Grant No. 2022YFA1402901), Science Fund for Distinguished Young Scholars of Shaanxi Province (No. 2024JC-JCQN-09), the Natural Science Foundation of Shaanxi Province (Grant No. 2023-JC-QN-0768), the Natural Science Foundation of China (Grant No. 12074301), and Science and technology plan of Xi'an  ( Nos.2021XDJH05, 202211080023). 
\end{acknowledgments}

\end{document}